\documentclass[twocol]{ametsocV6.1}

\usepackage[utf8]{inputenc} 
\usepackage[T1]{fontenc}    
\usepackage[english]{babel}
\usepackage[hidelinks]{hyperref}       
\usepackage{url}            
\usepackage{booktabs}       
\usepackage{amsfonts}       
\usepackage{nicefrac}       
\usepackage{lipsum}         
\usepackage{graphicx}
\usepackage{natbib}
\usepackage{doi}
\usepackage{multirow}
\usepackage{makecell}
\usepackage{gensymb}
\usepackage{dsfont}
\usepackage{amsmath}
\usepackage{bm}
\usepackage{float}
\usepackage{multicol}
\usepackage{setspace}
\usepackage{tabularx}
\usepackage{float}

\usepackage[acronym]{glossaries}
\usepackage{cleveref}       
\usepackage{wrapfig}

\title{STIPP: Space-time in situ postprocessing over the French Alps using proper scoring rules}

\authors{David Landry,\aff{a}\correspondingauthor{David Landry, david.landry@inria.fr} 
Isabelle Gouttevin,\aff{b} 
Hugo Merizen,\aff{b} 
Claire Monteleoni,\aff{a,c} 
Anastase Charantonis\aff{a}
}

\affiliation{\aff{a}{Inria, Paris, France}\\
\aff{b}{Météo-France, CNRS, Univ. Grenoble Alpes, Univ. Toulouse, CNRM, Centre d’Études de la Neige, 38000 Grenoble, France}\\
\aff{c}{University of Colorado Boulder, Boulder, United States}
}

\newacronym{AI}{AI}{Artificial Intelligence}
\newacronym{BS}{BS}{Brier Score}
\newacronym{CRPS}{CRPS}{Continuous Ranked Probability Score}
\newacronym{drn}{DRN}{Distributional Regression Network}
\newacronym{ecc}{ECC}{Ensemble Copula Coupling}
\newacronym{ES}{ES}{Energy Score}
\newacronym{esgm}{ESGM}{Energy Score Generative Model}
\newacronym{fmap}{FMAP}{Flow MAtching Postprocessing}
\newacronym{GAN}{GAN}{Generative Adversarial Network}
\newacronym{ode}{ODE}{Ordinary Differential Equation}
\newacronym{nudft}{NUDFT}{Non-Uniform Discrete Fourier Transform}
\newacronym{sde}{SDE}{Stochastic Differential Equation}
\newacronym{nlp}{NLP}{Natural Language Processing}
\newacronym{si}{SI}{Stochastic Interpolant}
\newacronym{RMSE}{RMSE}{Root Mean Squared Error}
\newacronym{VS}{VS}{Variogram Score}
\newacronym{lvs}{LVS}{Local Variogram Score}
\newacronym{fm}{FM}{Flow Matching}
\newacronym{fmt}{FMT}{Flow Matching Transformer}
\newacronym{NWP}{NWP}{Numerical Weather Prediction}
\newacronym{ifs}{IFS}{Integrated Forecasting System}
\newacronym{MLP}{MLP}{Multi-Layer Perceptron}
\newacronym{mbm}{MBM}{Member-by-member}
\newacronym{QRN}{QRN}{Quantile Regression Network}
\newacronym{est}{EST}{Energy Score Transformer}
\newacronym{SER}{SER}{Spread-Error Ratio}
\newacronym{scs}{ScS}{Schaake Shuffle}
\newacronym{silu}{SiLU}{Sigmoid Linear Unit}
\newacronym{ens}{ENS}{ECMWF Ensemble Prediction System}
\newacronym{EUPPBench}{EUPPBench}{EUMETNET Postprocessing Benchmark}
\newacronym{STIPP}{STIPP}{Space-time in situ probabilistic postprocessing}
\newacronym{GNN}{GNN}{Graph Neural Network}
\newacronym{PSD}{PSD}{Power Spectral Density}

\abstract{We propose Space-time in situ postprocessing (STIPP), a machine learning model that generates spatio-temporally consistent weather forecasts for a network of station locations.
Gridded forecasts from classical numerical weather prediction or data-driven models often lack the necessary precision due to unresolved local effects.
Typical statistical postprocessing methods correct these biases, but often degrade spatio-temporal correlation structures in doing so.
Recent works based on generative modeling successfully improve spatial correlation structures but have to forecast every lead time independently.
In contrast, STIPP makes joint spatio-temporal forecasts which have increased accuracy for surface temperature, wind, relative humidity and precipitation when compared to baseline methods.
It  makes hourly ensemble predictions given only a six-hourly deterministic forecast, blending the boundaries of postprocessing and temporal interpolation.
By leveraging a multivariate proper scoring rule for training, STIPP contributes to ongoing work data-driven atmospheric models supervised only with distribution marginals.} 

\begin{document}

\maketitle

\section{Introduction}

Gridded forecasts produced by \gls{NWP} models or their data-driven counterparts~\citep{BiPanguWeather3D2022,KurthFourCastNetAccelerating2023,PriceProbabilisticWeather2025} often lack in precision when compared against ground observations.
Some of these defects are simply explained, such as misalignment between the grid point and station elevation.
Others are not easily modeled, as they are caused by local effects that are ill-represented in the gridded model due to lack of resolution or inappropriate parameterization.
This is often addressed through a statistical correction step after the gridded forecast is computed, in a step called weather forecast postprocessing~\citep{VannitsemStatisticalPostprocessing2021}.

Postprocessing is a rich field of study within meteorology, providing a large array of tools tailored to different variables and time horizons.
However, this statistical calibration step sometimes has unwanted effects on the forecast, like the destruction of spatio-temporal correlation structures in the forecast fields.
Models with capabilities beyond marginal forecasts for each location and timestep are an ongoing area of research~\citep{FeikGraphNeural2024,ChenGenerativeMachine2024,LakeMultivariateEnsemble2025,LakatosCompositeLossGraph2025}.


In the context of an \gls{AI} revolution in weather forecasting~\citep{BouallegueRiseDataDriven2024}, new techniques are available to tackle these challenges.
Generative modeling, the family of \gls{AI} models encompassing \glspl{GAN}, diffusion and flow matching models~\citep{GoodfellowGenerativeAdversarial2014,HoDenoisingDiffusion2020,LipmanFlowMatching2023}, is well suited to preserve spatial and temporal correlation structures.
These methods output samples of the target distribution, rather than outputting an explicit representation of the distribution.
This allows modeling distribution with dimensionalities that were formerly intractable, at the cost of some interpretability.
In that spirit, generative modeling has already been applied to several problems in postprocessing, particularly where spatial structures are of interest~\citep{DaiSpatiallyCoherent2021,ChenGenerativeMachine2024,LandryGeneratingEnsembles2025}.

Recently, a new generative modeling technique has emerged for weather forecasting applications.
It consists in predicting the joint distribution of the atmospheric state by supervising only its marginal distributions~\citep{LangAIFSCRPSEnsemble2024,AletSkillfulJoint2025} using a proper scoring rule~\citep{GneitingStrictlyProper2007}.

Following this, we propose \gls{STIPP}, a probabilistic in situ weather forecasting model that represents both space and time correlation structures using a generative model trained using a proper scoring rule.
We demonstrate its predictive ability on a bespoke dataset covering the French Alps, making ensemble predictions for surface temperature, relative humidity, wind and precipitation.
\gls{STIPP} improves forecast accuracy when compared to baseline methods.
Its forecasts contain well-defined spatial structures, and have improved temporal continuity compared to generating forecasts independently at every time step.
Furthermore, \gls{STIPP} is designed to minimize constraints on the underlying \gls{NWP} forecast, ensuring it can flexibly be integrated to forecasting pipelines.
It  makes hourly ensemble predictions given only a six-hourly deterministic forecast, blending the boundaries of postprocessing and temporal interpolation.
This is relevant given the increased step size of data-driven weather forecasting models, which typically ranges between 6 and 24 hours~\citep{AletSkillfulJoint2025,CouaironArchesWeatherArchesWeatherGen2024}.
Finally, by successfully leveraging a multivariate proper scoring rule for its training, STIPP contributes to an ongoing body of work about generative model based on proper scoring rules in atmospheric sciences~\citep{LangAIFSCRPSEnsemble2024,AletSkillfulJoint2025,LarssonCRPSLAMRegional2025}.

This work is structured as follows.
Section \ref{sec:related-work} gives an overview of previous work related to probabilistic spatio-temporal in situ postprocessing.
Section \ref{sec:methods} describes the STIPP model in more details while Section \ref{sec:experiments} describes our experimental test bench, including dataset, model training and evaluation methods.
The results are depicted in Section \ref{sec:results} and discussed in Section \ref{sec:discussion}.
Section \ref{sec:conclusion} contains our concluding remarks.

\section{Related work}
\label{sec:related-work}

Our proposed model performs a combination of multivariate weather forecast postprocessing, time series forecasting, and weather forecast interpolation. 
While this is a novel combination to the best of our knowledge, it is related to ongoing works in all of these areas.

Multivariate ensemble forecast postprocessing is typically applied to the relation between different quantities at a same location, the relation between spatial locations, or both.
Frequently encountered methods involve a two-step process, where the forecast is at first postprocessed separately at the margins, before modifying the rank correlation structure to match a relevant dataset~\citep{ClarkSchaakeShuffle2004,SchefzikUncertaintyQuantification2013,BaranJointProbabilistic2015,LakatosComparisonMultivariate2023}.
An alternative is member-by-member postprocessing~\citep{SchaeybroeckEnsemblePostprocessing2015}, where an underlying ensemble is adjusted via scale and displacement factors independently for each variable, location and time step.
This has the advantage of preserving rank correlation structures, which yields consistent results in practice.
The main factor distinguishing STIPP from most of these methods is that it does not assume that an underlying ensemble forecast is provided, but only a deterministic one.
This loosens the constraints on the underlying forecast, at the cost of requiring more sophisticated postprocessing approaches.
The Schaake Shuffle~\citep{ClarkSchaakeShuffle2004} is compatible with this requirements and will be used as a baseline in our experiments.

On the time series forecasting side, previous work in solar irradiance involve neural network architectures closely related to ours, incorporating both gridded and in situ data through a transformer~\citep{BoussifImprovingDayahead2023,SchubnelSolarCrossFormerImproving2025}.
By predicting quantiles at every location, these models do make probabilistic forecasts but do not model correlation structures between neighboring sensors.

This work is closely related to applications of generative modeling in weather forecast postprocessing.
In an early adoption, \citet{DaiSpatiallyCoherent2021} use a \gls{GAN} to make cloud cover forecasts, a problem which requires a proper modeling of spatial structures.
More specifically related to in situ forecasting, \citet{ChenGenerativeMachine2024} propose a proper scoring rule generative model for temperature and wind forecasts.
This is done using an economical neural network architecture, which is extended to a \gls{GNN} by \citet{LakatosCompositeLossGraph2025}.
\citet{LandryGeneratingEnsembles2025} performs spatially-consistent postprocessing using a flow matching transformer.
These works perform time-independent forecasting, leaving a gap for spatio-temporal generative forecasting.

Recent developments in data-driven weather forecasting also informs our work in two ways. 
The first point of interest is the increased used of generation based on proper scoring rules~\citep{LangAIFSCRPSEnsemble2024,AletSkillfulJoint2025,LarssonCRPSLAMRegional2025}.
For gridded weather forecasting, these models provide more accurate forecasts while are significantly less expensive at inference.
This informs our choice to prioritize proper scoring rule generation over flow matching and diffusion approaches~\citep{PriceProbabilisticWeather2025,CouaironArchesWeatherArchesWeatherGen2024}.
The second point of interest is the observation that these models have significantly coarser temporal resolutions than their \gls{NWP} counterparts (6 hours by step or above), motivating research in temporal interpolation solutions such as the one proposed here.

\section{Methods}
\label{sec:methods}

This section introduces \gls{STIPP} progressively, first through a problem statement and training procedure, before introducing the components of its architecture (encoder, processor, decoder).
We also introduce two variants of this model, STIPP-CRPS and STIPP-TI.
The first is a simple modification of the training objective, while the latter is trained to make time-independent forecasts rather than outputting full time series at once.

\subsection{In situ postprocessing}

STIPP performs joint spatio-temporal in situ postprocessing based on temporally sparse NWP output.
This is framed as a supervised learning problem: a model is trained on past predictor-observations pairs so that it can infer future observations only from the predictors.
Here, the input predictors have two components: a gridded NWP forecast, provided every 6 hours for 42 hours, as well as 6 hours of recent observations at all station locations.
The model is tasked to predict time series of hourly observations at all spatial locations for all target variables.
This is done in a probabilistic way -- part of the input is randomized, and the model produces ensembles of timeseries, optimized for sharpness and reliability.

\begin{figure}
\centering
\includegraphics{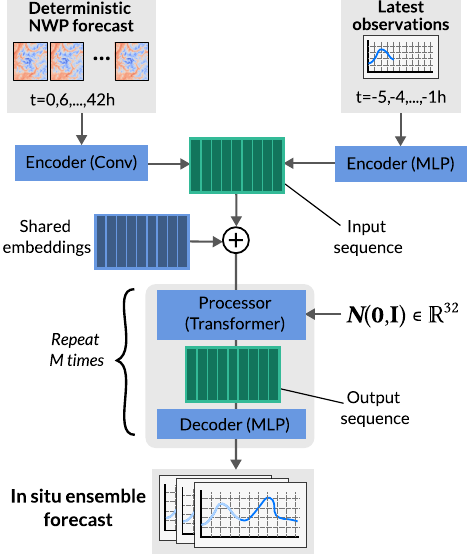}
\caption{
STIPP system diagram.
The model makes joint probabilistic time series forecasts for a set of station locations, given a gridded NWP forecast and recent observations at these locations.}
\label{fig:overview}
\end{figure}

This can be visualized on Figure~\ref{fig:overview}.
At the top, data from a deterministic NWP forecast and recent observations from the target locations are encoded into an input token sequence.
These two sequences are concatenated, and then their tokens are augmented with shared embeddings representing time and space information.
The composition of the input sequence is described thoroughly in Section \ref{ssec:encoder}.

On top of the input sequence, the processor also accepts low-dimensional random noise as input, which is sampled repeatedly to build an ensemble forecast.
This changing input allows the processor to be called multiple times to generate distinct ensemble members given one set of input predictors.

\subsection{Probabilistic training procedure}
\label{ssec:probabilistic-training}

STIPP generates ensembles through a proper scoring rule training procedure, which follows recent work in atmospheric sciences~\citep{ChenGenerativeMachine2024,LangAIFSCRPSEnsemble2024,AletSkillfulJoint2025,LarssonCRPSLAMRegional2025}.
Since proper scoring rules reward both sharpness and reliability, the model learns to be sensitive to its noise input, and learns to generate spread where appropriate.
Firstly, stochasticity is injected into the model by giving low-dimensional random noise as an input feature to the processor.
The exact injection mechanism is described in section~\ref{ssec:processor}.
Secondly, the model is called multiple times at every training step, so that an $M$-sized ensemble is created.
The value of $M$ is chosen as a hyperparameter, though in practice it is constrained by the GPUs capacity to hold multiple forward passes in memory.
This ensemble is evaluated against the corresponding observation using the chosen scoring rule.

We consider two such scoring rules, the \gls{CRPS} and its multivariate equivalent the \gls{ES}~\citep{GneitingStrictlyProper2007}.
Intuitively, we would expect that the multivariate \gls{ES} is preferable, because it is sensitive to the correlation structures between variables.
There are no fundamental guarantees that supervising a model only on the marginal distributions with the CRPS should be able to capture such representations.
However, there is mounting evidence that supervision on marginals achieves good performance on weather forecasting~\citep{LangAIFSCRPSEnsemble2024,AletSkillfulJoint2025,LarssonCRPSLAMRegional2025}.
These models systematically outperformed more theoretically-grounded counterparts such as diffusion and flow matching models, despite supervising only on marginals.
Furthermore, it is doubtful that the Euclidian norm in the \gls{ES} would be sensitive enough to pick up all relevant structures in a full forecast, given the extremely high dimensionalities involved.
This would amount to describing very high dimensional distributions using very small ensembles.

Several factors distinguish our problem from global weather forecasting which may invite for different strategies.
Our in situ forecasting problem dimensionality is much smaller (>100 000 dimensions in our case).
Furthermore, our model forecasts the temporal dimension directly, whereas previously mentioned models typically are auto-regressive.

Given this, we perform experiments using two training losses.
Our baseline STIPP model will use the \gls{ES} computed on the time step dimension.
Given $\bm{x}^{1..M}$ the  $T$-long generated time series for one station-variable pair, and $\bm{y}$ the corresponding observations, the loss is
\begin{align}
    \mathcal{L}_{\text{ES}} =  \frac{1}{M} \sum_{i=1}^M \| \bm{x}^i - \bm{y} \| - \frac{1}{2M(M-1)} \sum_{i,j=1}^M  \|\bm{x}^i - \bm{x}^j \|,
\end{align}
with $\|\cdot\|$ the euclidian norm.
This is averaged for all station-variable combinations in a training example.
To avoid an imbalance between variables operating at different scales, these losses are computed on normalized fields, with the normalization procedure stated in Section~\ref{ssec:normalization}.

We also perform experiments with a marginal loss defined as
\begin{align}
    \mathcal{L}_{\text{CRPS}} = \frac{1}{T}  \sum_{t=1}^T  \biggl[ \frac{1}{M}&\sum_{i=1}^M | x_t^i - y_t | - \frac{1}{2M(M-1)} \sum_{i,j=1}^M | x_t^i - x_t^j | \biggl],
    \label{eqn:loss-crps}
\end{align}
where $x_t$ and $y_t$ are respectively the $t$-th timesteps of $\bm{x}$ and $\bm{y}$, while $|\cdot|$ is the absolute value.
These two losses use the ``fair'' version of their respective scoring rules, which accounts for empirical ensemble size~\citep{FerroFairScores2014}.

\subsection{Model architecture}

\begin{figure*}[t!]
\includegraphics[width=\textwidth]{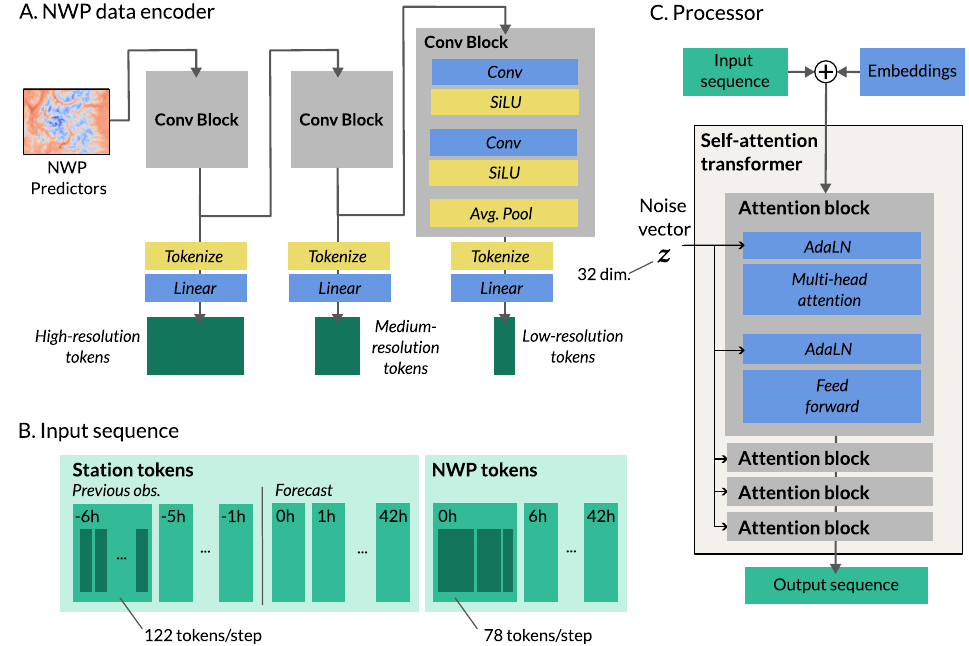}
\caption{The STIPP architecture. 
\textbf{Top left}. The NWP data encoder is used in conjunction with the station data encoder (not depicted) to build the input sequence. 
\textbf{Bottom left}. The input sequence contains tokens representing past and future states for each station, as well as tokens representing the NWP forecast.
The forecast station tokens are placeholders which will be populated by the processor.
\textbf{Right}. The processor is a self-attention transformer.
It treats the encoded token sequences after relevant spatial, temporal and semantic embeddings are added.
Stochasticity is injected through its conditional layer norm modules to 
enable generative forecasting.}
\label{fig:architecture}
\end{figure*}

\label{ssec:architecture}

We introduce the STIPP architecture in three sections, namely and encoder, processor and decoder.
Its key components are depicted in Figure~\ref{fig:architecture}.

\subsubsection{Encoder}
\label{ssec:encoder}

The purpose of the encoder is to build a sequence of $d$-dimensional tokens the represents the problem well, and that will allow the transformer to build relevant representations through attention mechanisms. 
It has three main components: a gridded forecast encoder, an in situ data encoder, and spatio-temporal embeddings.

The gridded data encoder is largely inspired by the U-Net architecture~\citep{RonnebergerUNetConvolutional2015}, although we only use its encoder side. 
The gridded data is progressively projected to lower resolutions, but with deeper embedding sizes.
This is depicted in Figure~\ref{fig:architecture}a.
The sequence of three convolutional blocks outputs latent representation sizes of $d/4$, $d/2$ and $d$, respectively.
At the output of each block, the encoder performs tokenization to represent the grid at different spatial resolutions.
To do so the grid is separated in square patches, as is typical with visual transformers~\citep{DosovitskiyImageWorth2021}.
Each patch is summarized through mean pooling, then projected with a linear layer to increase its embedding size to $d$.
This results in a sequence of tokens representing the gridded forecast at a high, medium and low resolution, respectively containing 58, 16 and 4 tokens.
Given each Arome forecast provides 8 time steps, our input sequences contain 608 tokens stemming from the gridded forecast.
This is concatenated to the input sequence represented in Figure~\ref{fig:architecture}b.

The in situ segment of the input sequence contains two types of tokens to represent the recent observations and the forecast. 
The past observation tokens are built using a \gls{MLP} with three hidden layers to encode the observation into size $d$. 
The forecast tokens use a separate \gls{MLP}, encoding the most recent observations for that station, as well as the NWP forecast value for the predicted variables, interpolated linearly in time.
The rationale is to fill this token with the most relevant information for either a nowcasting-like forecast or a NWP-based one.
This information is expected to be processed and modified by the transformer through attention to the other tokens and and the feed-forward layers.
Given 6 hours of previous observations, 42 lead times and 122 stations, this results in 5856 tokens for our specific instance.

Before the tokens are transmitted to the processor, they are augmented with relevant embeddings.
These embeddings play a role similar to the positional embedding in the original transformer~\citep{VaswaniAttentionAll2017}, that is, provide them with a representation of their relative position in space and time.
We use two such embeddings: a shared spatio-temporal embedding, and a station-specific embedding.
The shared spatio-temporal embedding is a \glspl{MLP} with two hidden layers.
It accepts latitude, longitude and elevation as input, time-of-day, day-of-year and lead time as inputs.
This embedding is used to add information to both gridded and in situ tokens.
The mean values of latitude, longitude and elevation are used to compute the NWP token embeddings.
The intuition behind sharing the space and time embeddings between gridded and in situ data is that it should trigger cross-attention between nearby objects across modalities.
The station-specific embedding is meant to represent local effects that dictate the mapping between nearest NWP gridpoint and station observations.
Such embeddings have a long usage history for in situ postprocessing~\citep{RaspNeuralNetworks2018}.

\subsubsection{Processor}
\label{ssec:processor}

The sequence constructed in Section \ref{ssec:encoder} is fed to a transformer~\citep{VaswaniAttentionAll2017}.
It has a decoder-only architecture, meaning attention is free to spread between all pairs of tokens in the sequence.
Because this self-attention spreads the whole spatio-temporal sequence, the transformer is able to freely transfer relevant information between the gridded forecast, the past observation records, the station forecasts, across steps.

For the purpose of layer normalization, we use AdaLN~\citep{GuoAdaLNVision2022} conditional layer normalization modules.
They are conditioned by a low-dimensional vector of standard Normal latents $\bm{z} \sim \mathcal{N}(\bm{0},\bm{I})$ which is sampled before each generation.
This noise having a small dimensionality is hypothesized to be critical for properly representing the joint distribution, despite supervising the training only on marginals~\citep{AletSkillfulJoint2025}.
Too many degrees of freedom could allow the model to converge towards less structured solutions that could damage spatial consistency.

\subsubsection{Decoder}

The full sequence is processed throughout the transformer.
Before decoding, the tokens related to past observations and gridded forecast are discarded, such that only the tokens related to in situ forecasts remain.
The decoder itself is a \gls{MLP} with 3 hidden layers, which brings the tokens down to the dimensionality of our forecasts (4 values per station-hour in this case).

\subsection{Model variants}

We introduce two variants of STIPP for the purpose of performing comparative studies.
The first variant, STIPP-TI, is trained independently for each time step.
It shares its architecture with the mainline model, but its sequence only contains in situ forecast tokens representing one forecast step, rather than the full timeseries.
This results in significantly smaller sequence lengths during training, but the number of model calls is scaled accordingly at inference time.
STIPP-TI allows us to study the benefits of training jointly for all time steps.

The second variant is STIPP-CRPS, which uses the training loss stated at Equation~\ref{eqn:loss-crps}.
It is used in ablation studies related to the effect of using the Energy Score for training.

\section{Experiments}
\label{sec:experiments}

We perform experiments with the STIPP architecture on a dataset centered around the French Alps.
The target area provides a compelling in situ forecasting problem, where a rich topography is expected to introduce local effects that would be challenging to model without in situ postprocessing strategies.
The forecasting task is to make 42 hours forecasts for 2m surface temperature, 2m relative humidity, 10m wind speed and surface precipitation given a deterministic NWP forecast and 6 hours of previous observations.

\subsection{Dataset}
The study area is an Alpine domain (between 5.0°E and 8.1°E in longitude; and between 43.925°N and 46.4°N in latitude) and the study period ranges from January 2019 to October 2024.
Runs before 2024 are used for training and validation.
Forecasts initialized on the 26th of each month or above are used for validation.
Data from 2024 acts as our testing set.

We make use of analyses and forecasts from the AROME-France NWP system \citep{SeityAROMEFranceConvectiveScale2011}, that runs operationally over a large domain centered on metropolitan France.

For training, the variables from Table \ref{Table:AromeData} are extracted from the 00h UTC, 6h UTC, 12h UTC, 18hUTC operational runs for lead times from 0 to 42h. 0h lead time refers to an analysis while other lead times correspond to forecasts. For traditionally cumulative variables over time like precipitation or radiation, the given value at hour h corresponds to the mean flux between h-1 and h. 

\begin{table*}[h]
\caption{Forecast and analysis data from the AROME NWP system used in the present study. Cumulative variables are denoted by a *.}\label{Table:AromeData}
\footnotesize
\begin{center}
\begin{tabular}{ccc}
\toprule
\textbf{Short name} & \textbf{Long name} & \textbf{Level above ground}\\
\midrule
T& Air temperature & 2~m, 50~m, 500~m, 500~hPa, 850~hPa\\
T\_{sol} & Surface temperature & surface\\
Tmax & Maximum air temperature over the preceding hour & 2~m\\
Hu & Air relative humidity & 2~m, 10~m, 500~hPa, 850~hPa\\
U and V & Zonal and meridional components of the wind speed & 10~m, 100~m, 500~m, 500~hPa, 850~hPa \\
U\_{raf} and V\_{raf} & Zonal and meridional components of wind gust speed & 10~m\\
Precip* & Total precipitation (cumul of liquid, solid and graupel) & surface\\
LWnet\_d* & Downward net infrared radiation & surface\\
SWnet\_d* & Downward net solar radiation & surface\\
SW\_d* & Downward solar radiation & surface\\
SW$_{dir}$\_d* & Downward direct solar radiation & surface\\
LE* & Latent heat flux & surface\\
P & Atmospheric pressure & surface\\
TKE & Turbulent kinetic energy & 100~m\\
NEBBAS & Low-altitude cloud cover (< 2500~m above ground)& - \\
NEBHAU & High-altitude cloud cover (> 5000~m above ground)& -\\
Z & Geopotential & 850~hPa\\
Altitude & Grid-cell altitude & surface\\
\bottomrule
\end{tabular}
\end{center}
\end{table*}

Learning and evaluation data also comprise measurements from surface observation stations operated or used by Météo-France. We selected stations from the Alpine domain which have co-located measurements for the four variables of interest (2~m air temperature, 2~m relative humidity, 10~m wind speed and precipitation) over 90\% of the study period, yielding 122 locations.

\subsection{Data normalization}
\label{ssec:normalization}

For data normalization purposes, the variables predicted by our model get specific treatment before being used in training.
10m wind speed and precipitation are mapped to their logarithm $\log(1+x)$ to improve normality.
Relative humidity is expressed as a value within $[0,1]$.
The 2m temperature values are rescaled station-wise according to their mean and standard deviation in the training observation record.
The predictors extracted from the NWP output are rescaled in bulk using their mean and standard deviation across all gridpoints in the training set.
Log-scaled predicted variables (wind speed and precipitation) are represented twice in the predictors, with and without the scaling operation.

\subsection{Baselines}

We use two baseline comparisons for this work.
The first is the unprocessed output from the deterministic Arome model which is used as input for all methods.
Predictions are extracted for each stations using the nearest gridpoint.

The second is a \gls{QRN} based on a \gls{MLP} with three hidden layers.
In a first step, this baseline method makes separate statistical correction for all variables.
A \gls{MLP} with three hidden layers is trained for this task by optimizing the \gls{CRPS}~\citep{BrockerEvaluatingRaw2012}.
To make forecasts between the validity times of the six-hourly gridded predictors, we interpolate all predictors linearly in time.

The \gls{QRN} predicts quantiles for every station--time-step, which are then reorganized using a Schaake Shuffle procedure~\citep{ClarkSchaakeShuffle2004}.
The reference correlation structures are drawn from the observations of the training data.

Series of observations which start at the same hour and are valid within five days of the target initialization time are considered.
Given that our training set spans from 2019 to 2023, this yields approximately 44 candidates for every forecast, from which we draw the ensemble ensemble.
We note that precipitation forecasting would likely benefit from a more elaborate sampling strategy to isolate relevant precipitation events, tough this modeling choice was judged sufficient for baseline purposes.

\subsection{Model implementation and training}

The processor transformer contains four transformer blocks with four attention heads each.
It has an internal embedding size $d$ of 512.
The feedforward module increases the embedding size to 1024 before reducing it to 512 again.
The conditional layer norm modules accept normal samples $\bm{z} \in \mathbb{R}^{32}$.

The station data encoder and shared space-time encoder are implemented as \glspl{MLP} with respectively 3 and 2 hidden layers, using SiLU activations~\citep{ElfwingSigmoidWeightedLinear2017}.

\gls{STIPP} is trained in two phases.
The first step is a deterministic training, where it is optimized against an \gls{RMSE} training loss. 
During that phase, no stochasticity is injected into the network, and a usual layer norm is used inside the attention blocks of the transformer.
This is done for 100 epochs with an AdamW optimizer~\citep{LoshchilovDecoupledWeight2019}, a OneCycle scheduler~\citep{SmithSuperConvergenceVery2018}, and $10^{-4}$ maximum learning rate.

In a second phase, conditional layer norm weights are initialized randomly, and stochasticity injection is enabled as described in Section~\ref{ssec:probabilistic-training}.
This is done using the same configuration for 50 epochs.
The training ensemble size $M$ is set to 2 throughout.

\subsection{Evaluation}
\label{ssec:evaluation}

After training, we generate 16-member ensembles for every considered model over the testing set for evaluation purposes.
This size is considered sufficient given the reduced dimensionality of our problem with respect to gridded forecasting.

Our evaluation methodology for \gls{STIPP} comprises a variety of verification metrics in order to qualify several aspects of the generated forecast distributions.
We complement this with illustrative case studies to help build intuition about model behavior. 
We summarize our evaluation toolkit here, while precise definitions are provided in Appendix A where appropriate.

We first compute the \gls{RMSE} to acts as a sanity check of the considered models.
For evaluating ensemble models, it is computed against the forecast ensemble mean.
We then compute probabilistic evaluation metrics: the \gls{CRPS}, which is univariate, and the \gls{ES}, its multivariate equivalent.
Model dispersion is evaluated specifically with the \gls{SER}.

We use the \gls{VS} to evaluate the quality of certain correlation structures specifically.
It is used to measure the quality of the correlations spatially between stations given a variable.
The \gls{VS} has a poor signal to noise ratio when measured on weakly correlated quantities.
To avoid this, we measure spatial correlations only for small neighborhoods.
We compute the \gls{VS} on a vector of the values of the 5 nearest stations~\citep{ChenGenerativeMachine2024}.
This is done around all stations, separately per variable.

Since our stated goal is to forecast space-time distributions jointly, we wish to evaluate how close the temporal behavior of the models is to that of the observations.
This is done by computing the \gls{PSD} on the generated time series and comparing it to the observations.
A low value in the high frequencies indicates a forecast that changes slowly through time with respect to the observations, and vice versa.

For some figures, the values are depicted as a skill score, meaning that given a score $\bar{S}$ aggregated over the test examples, skill score $SS$ is 
\begin{align}
    SS = 1 - \frac{\bar{S}}{\bar{S}_{\text{baseline}}},
\end{align}
where $\bar{S}_{baseline}$ is the score of an appropriate baseline.

\section{Results}
\label{sec:results}

\subsection{Model skill and spatial correlation structures}

Table \ref{tab:metrics} provides performance metrics aggregated over all time steps for the considered models.
Only the \gls{VS} is provided for the Arome baseline, since it is a deterministic forecast.
All STIPP variants perform better than our \gls{QRN} baseline, illustrating the benefits of postprocessing jointly for all spatial locations for this application.
The time independent STIPP-TI has degraded performance with respect to the other variants, illustrating the benefits of a joint spatio-temporal training. 
STIPP and STIPP-CRPS perform similarly for the CRPS, but the mainline version outperforms the marginal variant on all variables for the Energy Score.
This suggests that providing supervision for temporal correlation structures slightly improved the quality of spatial correlations.

\begin{table*}
    \centering
    \caption{Performance metrics across the predicted variables for two baselines and three STIPP variants.
    AROME represents the raw forcast on which the other models are based, which involves a temporal interpolation between the provided steps.
    QRN-Schaake represents a quantile regression network reorganized with a Schaake Shuffle.
    STIPP is optimized using the Energy Score across the time step dimension.
    STIPP-CRPS is optimized using the CRPS.  
    STIPP-TI makes its predictions independently at every time step.
    For this evaluation the Energy Score is computed using the joint prediction for all stations.
    Results are averaged for all time steps.}
    \footnotesize
    \label{tab:metrics}
\setlength{\tabcolsep}{3.5pt}

\begin{tabular}{lcccccccccccc}
\toprule
 & \multicolumn{4}{c}{CRPS} & \multicolumn{4}{c}{Energy Score (Space)} & \multicolumn{4}{c}{Variogram Score} \\
  & \makecell{Temp.} & \makecell{Hum.} & \makecell{Wind} & \makecell{Precip.} & \makecell{Temp.} & \makecell{Hum.} & \makecell{Wind} & \makecell{Precip.} & \makecell{Temp.} & \makecell{Hum.} & \makecell{Wind} & \makecell{Precip.} \\
  \cmidrule(lr){2-5} \cmidrule(lr){6-9} \cmidrule(lr){10-13} 
AROME (Interpolated) & - & - & - & - & - & - & - & - & 1.38 & 3.08 & 4.65 & 2.99 \\
QRN-Schaake & 1.27 & 7.48 & 0.84 & 0.13 & 17.75 & 104.91 & 12.75 & 3.56 & 0.96 & 2.55 & 2.92 & 2.71 \\
STIPP & \textbf{0.76} & \textbf{4.74} & 0.63 & \textbf{ 0.10} & \textbf{10.70} & \textbf{69.18} & \textbf{9.46} & \textbf{2.81} & \textbf{0.45} & \textbf{1.51} & \textbf{2.33} & \textbf{2.03} \\
STIPP (CRPS) & 0.76 & 4.75 & \textbf{0.62} & \textbf{0.10} & 10.80 & 69.75 & 9.52 & 2.92 & 0.45 & 1.54 & 2.47 & 2.07 \\
STIPP (TI) & 0.78 & 4.91 & 0.63 & \textbf{0.10} & 11.20 & 72.65 & 9.81 & 2.95 & 0.48 & 1.68 & 2.56 & 2.07 \\
\bottomrule
\end{tabular}

\end{table*}

\begin{figure*}
\centering
\includegraphics[width=\textwidth]{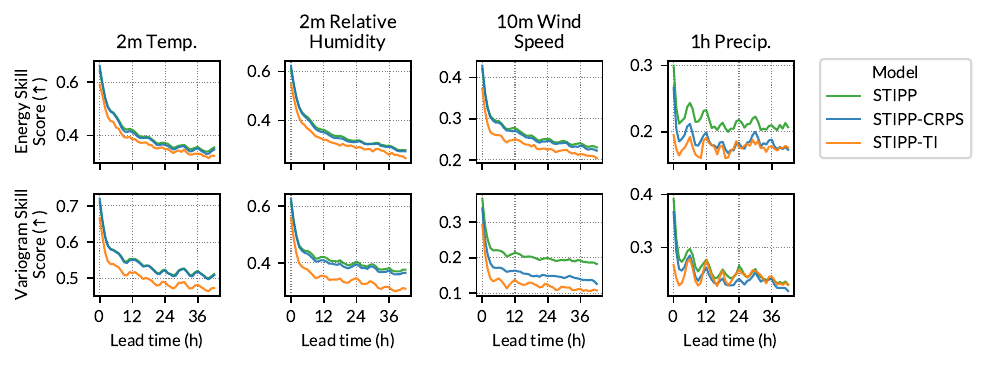}
\caption{
Skill scores for in situ weather forecasting according to lead time. 
The score are computed against the Schaake Shuffle baseline.
The upward arrow (↑) indicates that lower is better for all metrics.}
\label{fig:megaplot}
\end{figure*}

Figure \ref{fig:megaplot} shows Energy Skill Score and Variogram Skill Scores through lead time.
The skill scores are computed against the QRN-Schaake baseline.
All loss curves have a similar structure: the skill is higher for lead times under 6 hours.
We propose this indicates STIPP learned a nowcasting-like strategy for early lead times before relying on the NWP predictors, whereas the QRN baseline converged to only one postprocessing modality.
STIPP-TI has lower spatial variogram scores, indicating less ability to construct the local spatial correlation structures found in the observations.

\subsection{Power spectral density in time}
\label{ssec:power-spectrum}

\begin{figure*}
    \centering
    \includegraphics[width=\textwidth]{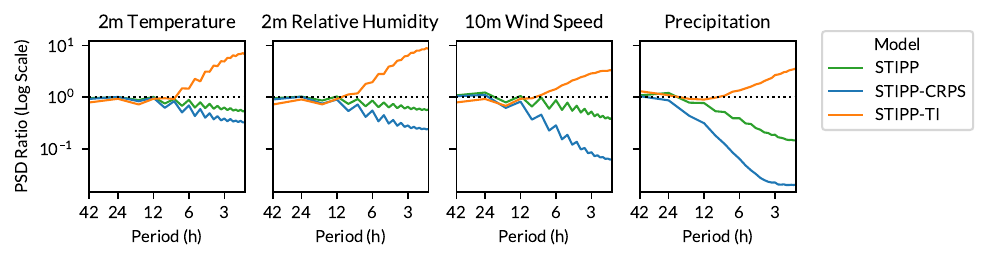}
    \caption{Power spectrum density of STIPP variants forecasts in the lead time dimension.
    The spectra are represented as a ratio over the spectrum of the observations.
    A value of one indicates the same energy signature as the observations for that period.}
    \label{fig:power-spectrum-time}
\end{figure*}

An important characteristic of STIPP with respect to previous work is the incorporation of the time dimension into in situ forecasting models.
Consequently, we pay a close attention to our models behavior in time.
Figure \ref{fig:power-spectrum-time} shows the power spectral density in time for the STIPP variants.
They are depicted as a ratio over that of the spectrum of the observations.

The time independent version has high power for shorter periods, which is unsurprising given that it makes forecasts independently for each time step -- no attempt is made to build time-continuous forecasts.
STIPP-CRPS has low power for shorter periods, indicating a lack of short-term variability in the forecasts.
The mainline STIPP has an intermediate power spectrum which is closer to the observations, but still lacks power in higher frequencies, especially for wind and precipitation.

\subsection{Dispersion}

\begin{figure*}
    \centering
    \includegraphics[width=\textwidth]{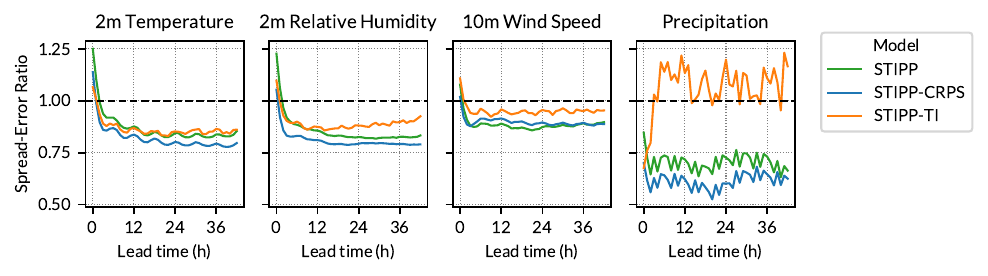}
    \caption{Spread-error ratios of STIPP variants.}
    \label{fig:spread-error-ratio}
\end{figure*}

We evaluate model dispersion using the spread-error ratio~\citep{FortinWhyShould2014}, depicted in Figures \ref{fig:spread-error-ratio}.
Some ratios are higher for early lead times, which is caused by the models having lower errors but similar dispersion at these steps.
Otherwise, the ratios for temperature, humidity and wind forecast indicate underdispersion for all model variants.
The underdispersion is more severe in precipitation for STIPP and STIPP-CRPS.
Interestingly, STIPP-TI is actually overdispersive for precipitation.
We posit that by making independent forecasts in time, it is better able to integrate an increased level of randomness in the forecasts, whereas the other models converged to more structured solutions.

\subsection{Precipitation forecasting}

We identify precipitation as a particularly interesting field for this study.
STIPP's spatio-temporal capabilities should enable it to approach precipitation forecasting in a way that is difficult using time and space independent methods.

\begin{figure*}
    \centering
    \includegraphics{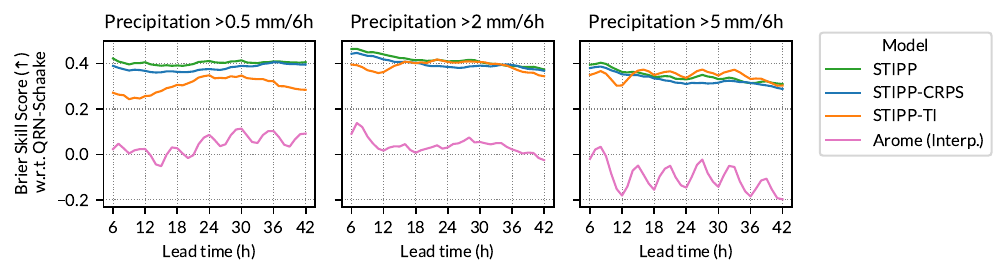}
    \caption{Model Brier Scores for predicting different precipitation thresholds accumulated over six hours. 
    The metrics are displayed as skill scores with respect to the QRN-Schaake baseline.}
    \label{fig:brier-score-precip}
\end{figure*}

In order to align with standard procedures of precipitation evaluation against station observations at Météo-France, we compute Brier scores for detection of 6 hours accumulation events.
We study three accumulation thresholds, corresponding to 0.5, 2 and 5 mm/h accumulated precipitation in 6 hours.
Figure \ref{fig:brier-score-precip} depicts this as a skill score against the QRN-Schaake baseline.
We consider that a precipitation is detected by the Arome baseline if the threshold is crossed within approximately 10 km (one grid point) of the station location, whereas the standard procedure at Météo-France dictates a threshold of 50 km.

The \gls{QRN} model with Schaake Shuffle fails to improve the raw deterministic forecast in the lower thresholds.
We attribute this to deficiencies in the sampling procedure from the observation record. 
Without efforts to pro-actively sample precipitation events, much of the suggested ranks would are random tie-breaks (due to most of the climatological samples having zero precipitation).
In contrast, STIPP methods are able to significantly increase the skill of the deterministic forecast without requiring a specific treatment for the precipitation variable.

\subsection{Case studies}

We plot two particular forecast instances from the test set which help build an understanding of our models behavior.

\begin{figure}
    \centering
    \includegraphics[width=\columnwidth]{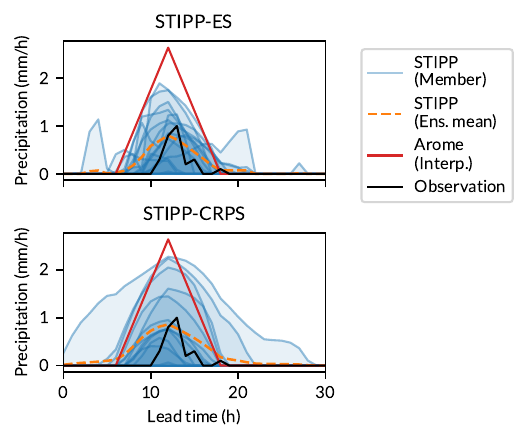}
    \caption{Example precipitation forecast for station Zermatt at initialization time 2024-01-03T12.
    Darker colors indicate more members forecast precipitation for a given lead time.}
    \label{fig:precipitation-timeseries}
\end{figure}

A sample precipitation forecast is visible in Figure \ref{fig:precipitation-timeseries} for the baseline STIPP and the STIPP-CRPS models.
The forecasts are valid for Zermatt station, initialized on 2024-01-03T12.
Both forecasts are somewhat successful in that the observed precipitation match ensemble means.
However, the shape of the forecasts is rather different, with STIPP-CRPS exhibiting very smooth and well ordered forecasts.
This suggests that the model builds forecasts in terms of statistical constructs, perhaps quantiles, rather than realizable ensemble members.
In contrast, the mainline STIPP forecasts have more variability in terms of temporal profiles, which is an improvement in terms of physical realizability.
This plot reflects different energy content in high frequencies already observed in Figure~\ref{fig:power-spectrum-time}.

\begin{figure*}
    \centering
    \includegraphics[width=\textwidth]{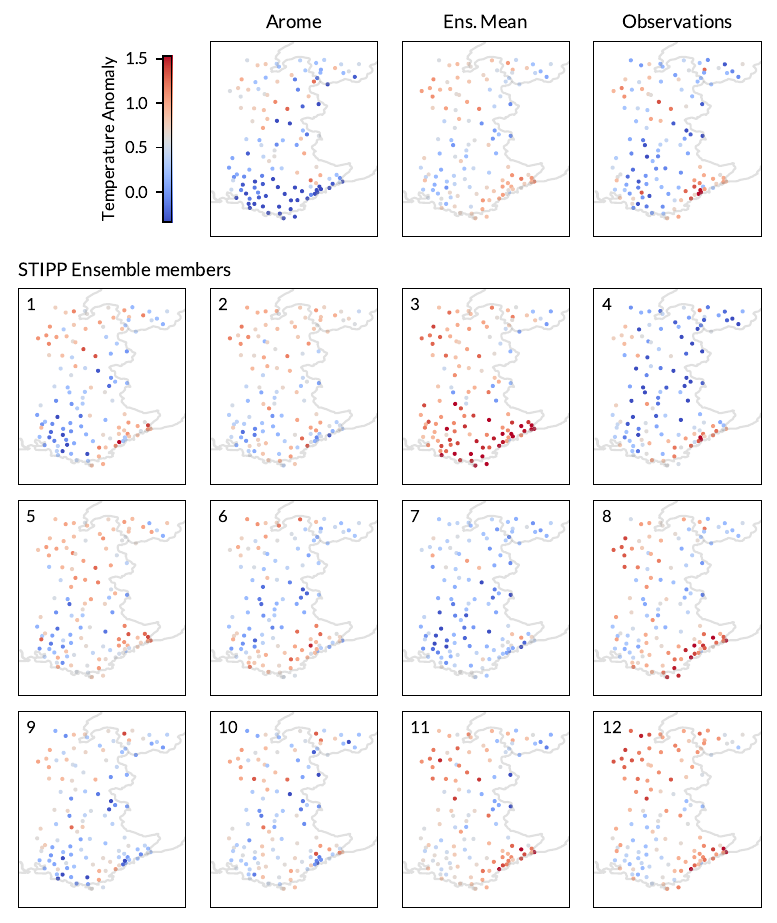}
    \caption{
    Example STIPP ensemble forecast of temperature anomaly.
    The top row displays the Arome forecast, the corresponding observation, as well as the STIPP ensemble mean.
    The following rows depict ensemble members generated using STIPP. 
    The first 12 members are depicted our of 16.
    Forecast initialization time: 2024-01-03T00. 
    Forecast lead time: 42h.
    }
    \label{fig:temperature-ensemble}
\end{figure*}

Finally, a sample STIPP ensemble temperature forecast is depicted in Figure \ref{fig:temperature-ensemble}.
The values are depicted as anomalies, according to a 21 day-of-year rolling-window climatology computed over the training set observation record.
The ensemble members correspond to spatially structured displacements of high temperatures.
Overall STIPP proposes a quite important corrections to AROME, especially in the South-Eastern region, in accordance to the observations.
Members 4 and 11 exhibit a structure similar to that of the observations, bringing the latter well within ensemble spread.

\section{Discussion}
\label{sec:discussion}

The results position STIPP as a skillful forecasting model despite using only a deterministic, temporally-sparse model as predictor.
All STIPP variants outperform a baseline trained independently in space and time.
The temporally-aware STIPP variants outperform STIPP-TI in most circumstances.
This underlines the benefits of our joint spatio-temporal approach.
STIPP has benefits beyond a typical postprocessing model, since it performs a more ambitious multivariate temporal interpolation task.
Rather, we see it as a demonstration of the flexibility brought by deep learning techniques for this type of forecasting tasks.

Among all the ablations performed, we find the difference in behavior between STIPP and STIPP-CRPS on Figure~\ref{fig:precipitation-timeseries} to be particularly striking.
In a context where training from marginals was successfully used in weather forecasting~\citep{LangAIFSCRPSEnsemble2024,AletSkillfulJoint2025,LarssonCRPSLAMRegional2025}, this work contributes to a better understanding of generative modeling through proper scoring rules.
We can suggest explanations as to why training only on marginals was sufficient in these works but was not ours.
For instance, these models are auto-regressive, which means they impose a Markovian structure to the models they converge to.
Our architecture imposes less structure, but in return demands more supervision from the loss function.
Still, we presented a counter-example where training only from marginals resulted in samples perceived as artificial.
This situation was partially improved by switching to the multivariate Energy Score training loss.
However, we do not believe this is a universal solution.
The length of our time series implied computing the Energy Score on 42-dimensional vectors, and it is doubtful its Euclidian norm  could efficiently reflect deficiencies in very high-dimensional distributions.
This suggests that choosing which variables to evaluate independently, and which variables to train on jointly, must be done on a case-by-case basis.
That question can be extended even further into the following ones: How robust are generative models supervised on marginals? Should their output be considered as samples from the distribution? If so, under which circumstances?

We anticipate future work related more specifically to forecasting at stations.
Our results suggest there is significant room for improvement in terms of high-frequency energy content of the forecasts, especially for wind and precipitation.
A better match with the observations would improve the physical realizability of STIPP forecasts.
The situation could be improved by introducing a variational component to the model, such that the forecast has an aleatoric component which we model as purely random.
Furthermore, the assumption that the injected noise vector $\bm{z}$ must be low-dimensional could be revised, seeing that the STIPP time series were not ``random enough'' for that purpose.
We also consider whether samples are the best forecasting product for such forecasts, or if well-defined statistics (e.g. mean, quantiles) are better suited for predicting values with high temporal variability.

Now that STIPP demonstrated flexibility by working from deterministic and temporally-sparse inputs, one can wonder how much improvement could be measured by changing the input to denser and probabilistic inputs. 
Also, we observe a convergence of methodologies between data-driven weather forecasting techniques and recent statistical correction methods.
As our postprocessing gains in complexity to address misspecifications in spatio-temporal correlation structures, the architectures we adopt gain in similarity to those of full weather forecasting models (i.e. by using a transformer, training on a proper scoring rule).
This raises the question of training the two systems jointly, rather than separating the weather forecasting and postprocessing trainings.
A potential benefit is the improvement of the gridded forecast through direct supervision from observations.
If this perspective is adopted, it follows that hybrid in situ-gridded models should be revisited~\citep{AndrychowiczDeepLearning2023a,YangLocalOffGrid2025}, with the hope that the models will learn more generalizable solutions if exposed to more of the atmospheric dynamics.

\section{Conclusion}
\label{sec:conclusion}

This work introduced STIPP, a probabilistic in situ spatio-temporal weather forecast postprocessing model.
Our model was able to provide skillful forecasts based only on a deterministic simulation, for four surface variables, on a challenging domain over the french Alps.
By operating on a temporally coarse input, STIPP demonstrated temporal interpolation capabilities that go beyond typical postprocessing models.
This fully spatio-temporal strategy proved relevant, improving forecast accuracy over an equivalent time-independent solution.
STIPP also showed the benefits of multi-variate training losses for proper scoring rule generative models, by improving the physical plausibility of forecasts compared to a model trained only on marginals.
This contribution opens new possibilities for in situ forecasting and raises interesting questions about training from proper scoring rules.
It also suggests future work avenues in hybrid gridded-in situ forecasting, with the view of blending boundaries between full weather forecasting and postprocessing.

\section*{Data availability statement}

The source code for the models and experiments can be obtained by contacting \url{david.landry@inria.fr}.
The Arome outputs and observations are subject to data sharing policies at Météo-France, please contact \url{hugo.merzisen@meteo.fr}.

\bibliographystyle{ametsocV6}
\bibliography{zotero_bibtex}

\onecolumn

\appendix[A]
\appendixtitle{Evaluation metrics}

\section{CRPS}

The \gls{CRPS}~\citep{GneitingStrictlyProper2007} is a proper scoring rule, meaning it is minimized when models predict the true distribution of observations.
This makes it a widespread tool in ensemble forecast verification. 
We approximate it using its \textit{fair} variant~\citep{FerroFairScores2014}:
\begin{align}
    \text{CRPS}(X_{t,k}, y_{t,k}) =\frac{1}{M}&\sum_{i=1}^M | x_{t,k}^i - y_{t,k} | - \frac{1}{2M(M-1)} \sum_{i,j=1}^M | x_{t,k}^i - x_{t,k}^j |.
\end{align}
This removes biases related to the size of the sample, so the value can be interpreted as an approximation of the \gls{CRPS} given an infinite ensemble size.
The \gls{CRPS} is a marginal verification tool, which is not sensitive to the correlation structures between forecast quantities.

\section{Energy Score}

The \gls{ES} is a multivariate extension of the \gls{CRPS}~\citep{GneitingStrictlyProper2007}, allowing us to evaluate both the marginal and multivariate quality of the distribution.
We compute it using
\begin{align}
    \text{ES}(\bm{X}_t, \bm{y}_t) = \frac{1}{M} \sum_{i=1}^M \| \bm{x}^i_t - \bm{y}_t \|^\beta - \frac{1}{2M(M-1)} \sum_{i,j=1}^M  \|\bm{x}^i_t - \bm{x}^j_t \|^\beta,
    \label{eqn:es}
\end{align}
using the fair variant here as well.
We use $\beta = 1$ so that it would reduce to the \gls{CRPS} in the one-dimensional case.

\section{Variogram Score}

The \gls{VS}~\citep{ScheuererVariogramBasedProper2015} is especially useful to quantify how accurately correlation structures are represented by our models.
It is defined as 
\begin{align}
    \text{VS}(\bm{X}_t, \bm{y}_t) =  \sum_{i,j=1}^N  \left( |y_{t,i} - y_{t,j}|^\rho - \frac{1}{M} \sum_{m=1}^M | x_{t,i}^m - x_{t,j}^m |^\rho \right)^2,
\end{align}
where we use $\rho=\frac{1}{2}$.
Since it is not sensitive to biases, it is especially useful when combined with the \gls{CRPS} or \gls{ES}, allowing us to evaluate the multivariate behavior specifically.

We compute the \gls{VS} in two ways within this work.
The first measures correlations across variables, for all station--lead-times.
The second measures the quality of the correlations spatially between stations.
The \gls{VS} has a poor signal to noise ratio when measured on weakly correlated quantities.
To avoid this, we measure spatial correlations only for small neighborhoods.
We compute the \gls{VS} on a vectors of the values of the 5 nearest stations~\citep{ChenGenerativeMachine2024}.
This is done around all stations, separately per variable.

\section{Spread-error ratio}

As a way to quantify the dispersion of the considered models, we compute the \gls{SER}~\citep{FortinWhyShould2014}, defined as 
\begin{align}
\text{SER} = \sqrt\frac{M+1}{M}\frac{\text{Spread}}{\text{Error}}    
\end{align}
with
\begin{align}
    \text{Spread} &= \sqrt{\frac{1}{T}\sum_{t=1}^T \frac{1}{M-1} \sum_{i=0}^M (x^i_{t,k} - \bar{x}_{t,k})^2} \\
    \text{Error} &= \sqrt{\frac{1}{T} \sum_{t=1}^T (\bar{x}_{t,k} - y_{t,k})^2}.
\end{align}
This ratio is expected to be close to one assuming exchangeability between the ensemble members, which prevents its use with quantile forecasts. 

\subsection{Brier Score}

The \gls{BS} is a proper scoring rule for ensemble forecasts predicting a binary event.
We use it to evaluate precipitation forecasts.
Given exceedance threshold $\tau$, we compute it using
\begin{align}
    \text{BS}_\tau(X_{t,k}, y_{t,k}) = \Big( \bm{1}[y_{t,k} > \tau] - \frac{1}{M} \sum_{i=0}^M \bm{1} [x^i_{t,k} > \tau] \Big)^2
\end{align}
where $\bm{1}[\cdot]$ is the indicator function.

\appendix[B]
\appendixtitle{Supplementary results}

\section{Rank histograms}

The figure below shows the rank histograms of 24h forecasts for STIPP models over the test set.

\begin{figure}[H]
    \centering
    \includegraphics[width=\textwidth]{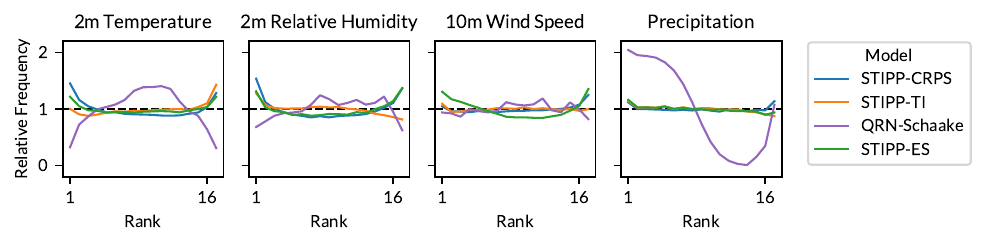}
    \caption{Rank histogram of STIPP models.}
    \label{fig:rank-histogram}
\end{figure}

\section{Precipitation forecast ensemble}

The figure below depicts a STIPP forecast for a high-intensity precipitation event observed on 2024-03-31T18. 
The underlying Arome deterministic forecast captures most of the intensity but is slightly spatially displaced, and does not forecast precipitation in the north-western part of our observation boundaries.
Several STIPP members depict events of lower intensity, though members 4 and 10 show good similarity with the observations.

\begin{figure}[H]
    \centering
    \includegraphics[width=\textwidth]{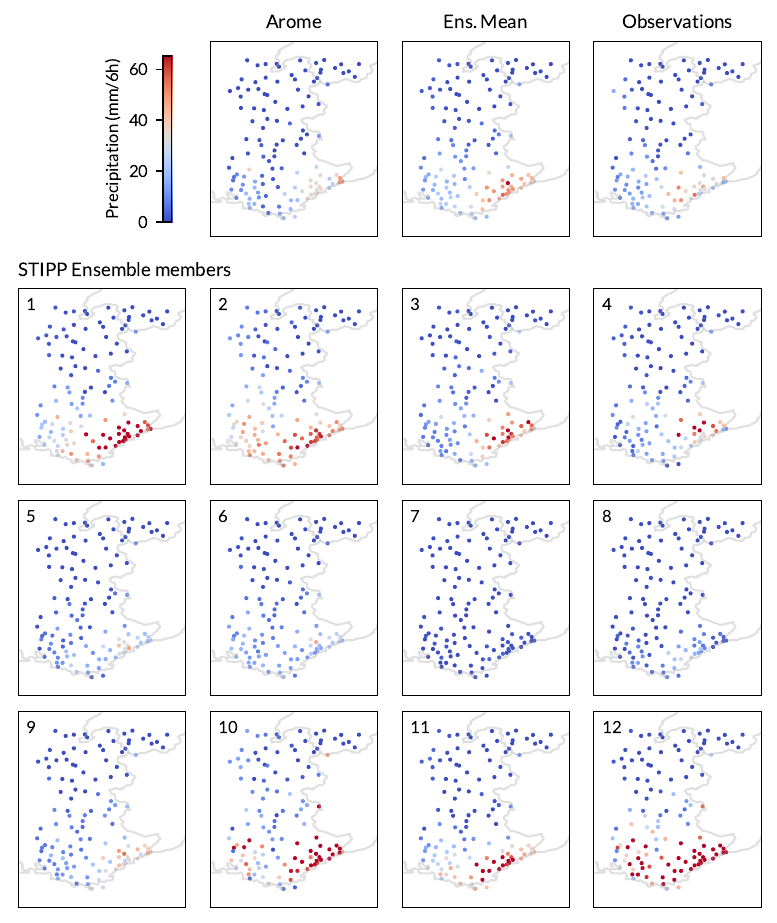}
    \caption{Example STIPP ensemble precipitation forecast.
    The top row displays the Arome forecast, the corresponding observation, as well as the STIPP ensemble mean.
    The following rows depict ensemble members generated using STIPP. 
    The first 12 members are depicted out of 32.
    Forecast initialization time: 2024-03-30T12. 
    Forecast lead time: 36h.}
    \label{fig:precip-acc}
\end{figure}

\section{Multivariate correlations}

\begin{figure}[H]
    \centering
    \includegraphics{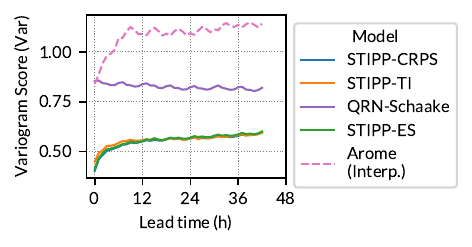}
    \caption{Variogram score for cross-variable correlations. The score is computed separately for each station at each time step.}
    \label{fig:variogram-score-var}
\end{figure}

The cross-variable correlation structures are evaluated using the \gls{VS} in Figure \ref{fig:variogram-score-var}.
Interestingly, the Arome forecast performs poorly, even on time steps \{ 6h, 12h, ... \} where a forecast is provided exactly.
This illustrate the difficulty of modeling local weather effects from a gridded forecast, even given a high-resolution deterministic forecast such as Arome.
The deterministic STIPP-D performs poorly as well, this time due to it predicting conditional expectations for every variable, blurring out correlations.
Our other models (STIPP, STIPP-TI, STIPP-CRPS) perform very similarly.

\end{document}